\documentclass[11pt,twoside]{article}

\usepackage{asp2006}
\usepackage{epsf}
\usepackage{psfig}
\usepackage{lscape}

\begin{document}
\title{{\it Spitzer}  Mid-Infrared Spectra of Cool-Core Galaxy Clusters}
\author{G. E. de Messi\`{e}res\altaffilmark{1}, 
        R. W. O'Connell\altaffilmark{1},
        B. R. McNamara\altaffilmark{2},
        M. Donahue\altaffilmark{3},
        P. E. J. Nulsen\altaffilmark{4},
        G. M. Voit\altaffilmark{3},
        M. W. Wise\altaffilmark{5}}

\altaffiltext{1}{P.O. Box 400325, University of Virginia, 
  Charlottesville, VA 22904-4325}
\altaffiltext{2}{Department of Physics \& Astronomy, University of Waterloo, 
  200 University Ave. W., Waterloo, Ontario, Canada N2L 3G1}
\altaffiltext{3}{Physics and Astronomy Department, 
  Michigan State University, East Lansing, Michigan 48824-2320}
\altaffiltext{4}{Harvard-Smithsonian Center for Astrophysics, 
  60 Garden Street, Cambridge, MA 02138}
\altaffiltext{5}{University of Amsterdam, Astronomical Inst Anton Pannekoek, 
  Kruislann 403, NL 1098 SJ Amsterdam, Netherlands}

\begin{abstract}

We have obtained mid-infrared spectra of nine cool-core galaxy clusters with 
the Infrared Spectrograph aboard the Spitzer Space Telescope. X-ray, 
ultraviolet and optical observations have demonstrated that each of these 
clusters hosts a cooling flow which seems to be fueling vigorous star formation
in the brightest cluster galaxy. Our goal is to use the advantages of the 
mid-infrared band to improve estimates of star formation. Our spectra are 
characterized by diverse morphologies ranging from classic starbursts to flat 
spectra with surprisingly weak dust features. Although most of our sample are 
known from optical/UV data to be active star-formers, they lack the expected 
strong mid-infrared continuum. Star formation may be proceeding in unusually 
dust-deficient circumgalactic environments such as the interface between the 
cooling flow and the relativistic jets from the active galactic nucleus. 
\end{abstract}


\section{Introduction}

Brightest cluster galaxies (BCGs) have elliptical morphologies and generally do
not host significant star formation.  However, a subset of BCGs are marked by 
their cuspy X-ray emission \citep{crawford99,edwards07} and a high incidence of
strong nebular line emission, which indicates star formation or nuclear 
activity.  The intracluster gas in these cool-core clusters is thought to be 
condensing in the form of a ``cooling flow'' \citep{fabian94}.  Many such 
clusters have blue continua and star formation rates (SFRs) as high as 
$\sim 100\,{\rm M}_{\odot}\,{\rm yr}^{-1}$, suggesting that the star formation 
is fueled by the cooling flows.

Recent X-ray spectroscopy has demonstrated that cooling flows are subject to 
regulatory heating, probably associated with nuclear activity 
\citep{mcnamara07}.  The estimated net rates at which gas cools below 1 keV are
comparable with the estimated star formation rates \citep{odea08, rafferty06}.
Cool-core clusters present a prime opportunity to test models of the cooling 
and energy feedback mechanisms that are thought to dominate early galaxy 
formation.

Radio observations of CO emission and near-infrared detections of the 
ro-vibrational transitions of H$_2$ have revealed large reservoirs of cool 
molecular gas in many cooling flows \citep{edge01, donahue00}.  Furthermore, 
far-infrared photometry with the {\it Spitzer} Space
Telescope shows that some BCGs in cool-core clusters have spectral energy 
distributions consistent with emission from cool dust ($T_d \la 100$) 
\citep{egami06, odea08, quillen08}. There are similarities between classic 
starbursts \citep{brandl06} and BCGs in cool-core clusters, but many questions
about star formation in the BCGs remain.

We use {\it Spitzer} to study nine BCGs in cool-core clusters which have been 
extensively observed in X-ray, optical and ultraviolet light.  With one
exception, our targets are confirmed star-formers with optical and UV 
continuum-based SFRs ranging up to $130\,{\rm M}_{\odot}\,{\rm yr}^{-1}$.   
The mid-infrared (MIR) band contains many diagnostics of star formation, 
including emission by small warm dust grains and complex emission by polycyclic
aromatic hydrocarbons (PAHs).  It also contains many narrow emission features, 
including atomic emission associated with nuclear activity and emission lines 
from molecular hydrogen in pure rotational states.  Our goal is to establish a 
better understanding of the cooling flow, feedback and star formation processes
in BCGs.

\section{Observations}

The spectra in Figure 1 were taken with the four low-resolution modes of 
{\it Spitzer's} Infrared Spectrograph under Program ID's 3384 and 20345.
Our targets are Abell 1068, Abell 1835, PKS0745-19, Hydra A, Abell 1795, 
Abell 2597, Abell 478, ZwCl 1370, and 2A0335+096.  We obtained sparse spectral 
maps of all BCGs, but we restrict our current discussion to a central 
extraction that yielded the best signal-to-noise.  After subtracting background
light and correcting to the restframe of each BCG, we renormalized each 
spectrum to the total light from the target, using the light profile along the 
SL1 slit and assuming circular symmetry.

\begin{figure}
\epsscale{0.8}
\plotone{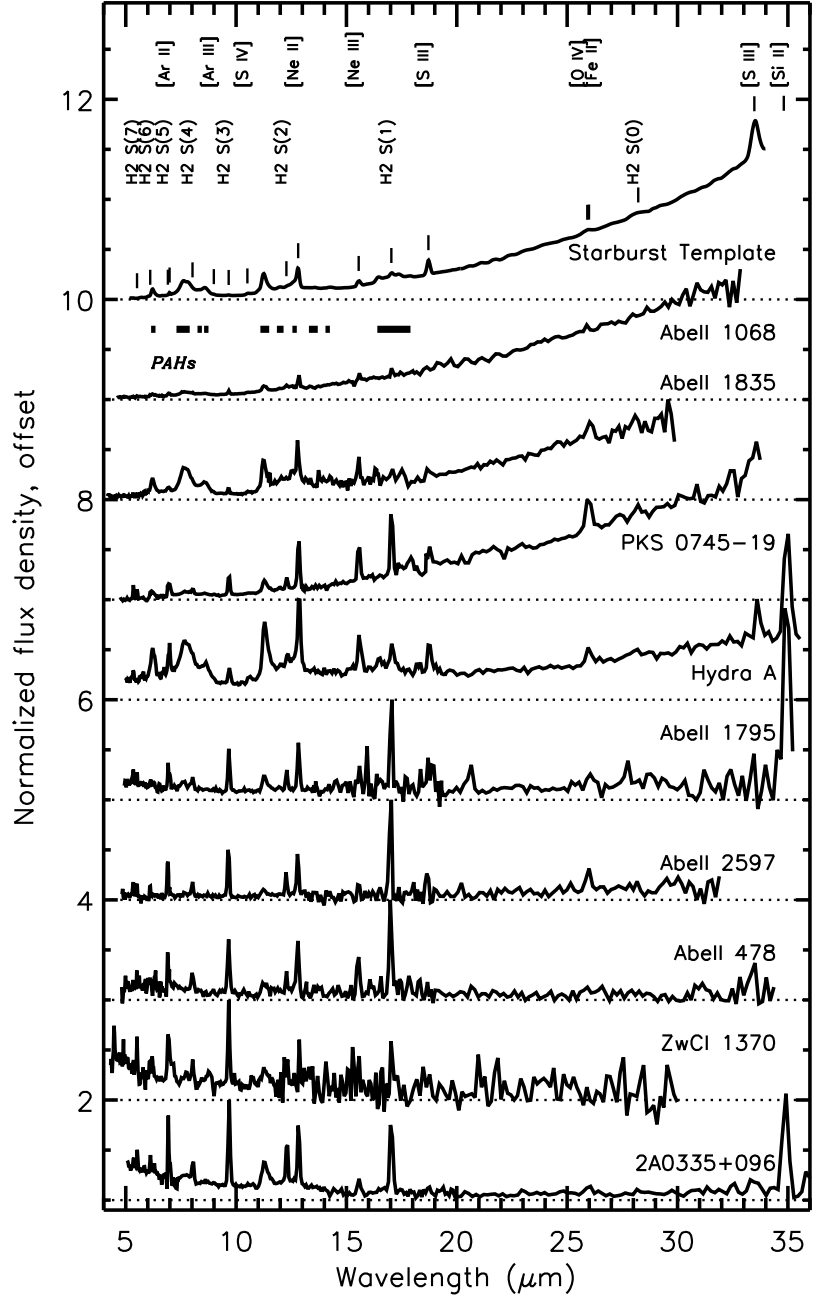}
\caption{IRS spectra of the nine BCGs, grouped by gross spectral morphology, 
and a starburst template spectrum \citep{brandl06}.  Spectra are plotted in
units of intensity per unit frequency (mJy/sr), normalized so that the highest 
point blueward of 30$\mu$m is equal to 1, and additively offset.  Unresolved 
emission lines and PAH bands are labeled at the top.  Note the strong emission
lines, particularly [Ne II] and the pure rotational series of H$_2$.  PAH 
emission is detected in all cases, but is weaker than normal in most.}
\end{figure}

\section{Discussion and Conclusion}

The resulting spectra are compared to a typical starburst template spectrum 
\citep{brandl06} in Figure 1.  Despite the high SFRs derived from optical and
UV data, most of our BCGs differ from normal starbursts in one or more of the 
following characteristics: weaker long-wave continua, weaker PAH features, and 
strong H$_2$ features.

The pure rotational lines of H$_2$ are abnormally strong in our spectra.  From
the line strengths, we infer a range of gas masses at different temperatures, 
similar to results discussed by \citet{ferland08}.  The strong H$_2$ features 
probably originate in cold gas reservoirs subject to shocks \citep{guillard08} 
and are probably unrelated to star formation.  We will discuss the quantitative
interpretation of the H$_2$ and other features in an ensuing paper.

Of our nine galaxies, only A1835 has the MIR spectral morphology of a classic 
starburst.  While two others exhibit the expected strong red continuum, they
also have anomalous emission feature spectra.  The remaining six spectra have 
weak red continua, despite a firm lower limit on star formation established for
five of them by optical and ultraviolet observations.  The well-established
correlations between IR output and star formation \citep{rieke09} break down.  
We have established that if the normal MIR continuum signatures of star 
formation were present in these galaxies, we would have easily detected them; 
instead, in most of our targets, the MIR continuum is subluminous by factors of
10-70x.

The abnormally weak MIR continua indicate anomalies in the amount of dust, in 
its grain size or composition, or in its spatial or temperature distribution.
Our results may indicate that star formation in BCGs in cool-core clusters is 
progressing in an unusual environment.  Rather than occurring in a dense, 
disk-like volume, as in normal starbursts, star formation may be concentrated 
in extended regions outside the main body of the galaxy, such as at the 
interface between relativistic jets from the AGN and the inner cooling flow.  
One example is NGC 1275, where star formation occurs in a huge filamentary 
network \citep{conselice01, fabian08, mcnamara96}.  In this unusual 
environment, the dust grain population is expected to differ considerably from 
the dust in normal starbursts and spiral disks.

\acknowledgements This work is based on observations made with the Spitzer 
Space Telescope, which is operated by the Jet Propulsion Laboratory, California
Institute of Technology under a contract with NASA. Support for this work was 
provided by NASA through two awards issued by JPL/Caltech.

\end{document}